\documentclass{PoS}
\pdfoutput=1
\usepackage{defs}
\usepackage{amsmath,mathrsfs,cite,booktabs}
\usepackage{amsfonts}

\def\LALPHA{
\hbox{\includegraphics[width=2.5cm]{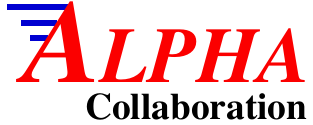}\hspace*{2cm}}
}
\newcommand{\onecol}[2]{
        \begin{minipage}[t]{#1}{#2\vfill} \end{minipage}
        }

\title{$M_{\rm b}$ and $\fB$ from non-perturbatively renormalized HQET with $\nf=2$ light quarks}

\ShortTitle{$M_{\rm b}$ and $\fB$ from NP'ly renormalized HQET with $\nf=2$ light quarks}

\author{\LALPHA \hfill
        \onecol{4.0cm}{\vspace{-1.5cm}
        \it  CERN-PH-TH/2011-318 \\
             DESY 11-251 \\ 
             Edinburgh 2011/40 \\ 
             HU-EP-11/63       \\
             LPT Orsay/11-127  \\
             MKPH-T-11-26      \\ 
             MS-TP-11-25       \\ 
             SFB/CPP-11-81     \\
        }}

\author{%
Beno{\^\i}t Blossier$^a$, 
John Bulava$^b$, 
Michele Della Morte$^c$,  
Michael Donnellan$^d$, 
\speaker{Patrick Fritzsch}\,$^e$, 
Nicolas Garron$^f$,  
Jochen Heitger$^g$, 
Georg von Hippel$^c$,
Hubert Simma$^d$, 
Rainer Sommer$^d$ \vspace{0.4cm} \\
\llap{$^a$} Laboratoire de Physique Th\'eorique, CNRS et Universit\'e Paris-Sud XI, 
B\^atiment 210, 91405~Orsay~Cedex, France\\
\llap{$^b$} CERN, Physics Department, CH-1211 Geneva 23, Switzerland \\
\llap{$^c$} Universit\"at~Mainz, Institut~f\"ur~Kernphysik, Becherweg~45, 55099~Mainz, Germany \\
\llap{$^d$} NIC, DESY, Platanenallee 6, 15738 Zeuthen, Germany \\
\llap{$^e$} Humboldt Universit\"at, Institut f\"ur Physik, Newtonstr. 15, 12489~Berlin, Germany \\
\llap{$^f$} Tait Institute, University of Edinburgh, Edinburgh EH9 3JZ, UK\\ 
\llap{$^g$} Universit\"at M\"unster, Institut f\"ur Theoretische Physik, Wilhelm-Klemm-Str. 9, 48149 M\"unster, Germany \\
}

\abstract{
We present an updated analysis of the non-perturbatively renormalized b-quark 
mass and $B$ meson decay constant based on CLS lattices with two dynamical 
non-perturbatively improved Wilson quarks. This update incorporates additional light 
quark masses and lattice spacings in large physical volume to improve chiral 
extrapolations and to reach the continuum limit. We use Heavy Quark Effective Theory 
(HQET) including $1/\mb$ terms with non-perturbative coefficients based on the matching of 
QCD and HQET developed by the ALPHA collaboration during the past years.
}

\FullConference{The XXIX International Symposium on Lattice Field Theory - Lattice 2011\\
July 10-16, 2011\\
Squaw Valley, Lake Tahoe, California}

\begin{document}

\section{Introduction}

In the current area of flavor physics, LHC's 2011 data taking pushes
the amount of available experimental data in $B$ physics to a new level
and expands the window of precision tests of the Standard Model. 
Due to confinement the processes considered at the experiments must involve
hadronic initial states. The impact on the indirect search of New Physics is 
therefore limited by the uncertainties on long distance effects.
A natural way to estimate these non-perturbative hadronic contributions within a few
percent of theoretical error is given by Lattice QCD. To carefully treat $B$ physics on the lattice 
one has to keep under control simultaneously the finite size effects and, particularly, the discretisation 
effects (the lattice spacing should  be smaller than the Compton wavelength of the b-quark) induced by the 
simulation. In practice it is not possible to control both effects in one simulation. 

The ALPHA Collaboration has followed a strategy discussed in detail in \cite{HeitgerNJ}: 
in HQET the hard degrees of freedom $\sim\, m_{\rm b}$ are removed and taken into 
account by an expansion in the inverse b-quark mass $m_{\rm b}^{-1}$. As discussed in those 
papers and also in earlier work, the benefit is the suppression of large discretisation 
effects which may arise in hadronic quantities when the theory is regularised on the lattice.
The difficult aspect of that method is that a matching with QCD, which is the field theory 
believed to describe the strong interactions, is needed to fix the parameters of the effective theory.
This step also takes care of removing all UV divergencies appearing in the effective theory. 
In lattice HQET those come as inverse powers of the lattice spacing and thus have to be removed 
non-perturbatively before the continuum limit can be taken. 

The main advantages of this method compared to other existing ones are: i) the
theoretical soundness of the approach, in particular the existence of a
continuum limit which is in addition numerically reachable, ii) the completely
non-perturbative treatment at any order in the $1/m_{\rm b}$ expansion,
including the matching between QCD and HQET, iii) the self-consistency of the
method, meaning that HQET is tested rather than assumed as opposed to what
other approaches have to do for masses around the charm, iv) the numerical
cost, which is comparable to that for other setups, as the extra computations
needed for the small-volume matching between QCD and HQET are very cheap
compared to large-volume simulations.

\section{Parameters and observables of Heavy Quark Effective Theory}
The HQET Lagrangian including terms of order $1/m_{\rm b}$ reads
\begin{align} \label{e:HQET-Lagrangian}
   \lag{HQET}(x) &=  \lag{stat}(x) - \omegakin\Okin(x)  - \omegaspin\Ospin(x)  \,,\, \\[0.125em]
\lag{stat}(x) &= \heavyb(x) \,D_0\, \heavy(x) \;,\quad
     \Okin(x) = \heavyb(x){\bf D}^2\heavy(x) \,,\quad 
    \Ospin(x) = \heavyb(x){\boldsymbol\sigma}\!\cdot\!{\bf B}\heavy(x)\,,
\end{align}
with the lowest order (static) Lagrangian $\lag{stat}$ 
and the first order $(1/m_{\rm b})$--corrections $\Okin$, $\Ospin$,
giving the kinetic and spin contribution respectively. These are sufficient to
compute the b-quark mass to subleading order, but in order to compute the
pseudo-scalar decay constant we also introduce the zero momentum 
projected time component of the heavy-light axial vector current,
\begin{align}
 \Ahqet(x_0) &= \zahqet\, a^3{\sum}_{{\bf x}}\,[\Astat(x)+  \cah{1}\Ah{1}(x)]\,, \\[-0.125em]
 \Astat(x)   &= \lightb(x)\gamma_0\gamma_5\heavy(x)\,,\qquad \qquad \qquad 
 \Ah{1}(x)    = \lightb(x)\frac{1}{2}
            \gamma_5\gamma_i(\nabsym{i}-\lnabsym{i}\;)\heavy(x)\,,
 \label{e:dahqet}
\end{align}
where $\nabsym{i}$ denotes the spatial components $(i=1,2,3)$ of the symmetric covariant derivative. 
The effective parameters of HQET that need to be known hence are 
\begin{align}
        \omega &= \big( \mhbare, \ln[\zahqet], \cah{1}, \omegakin, \omegaspin \big)\,.
        \label{eq:omegas}
\end{align}
The additional parameter $\mhbare$ in the static theory, which does not appear 
in~\eqref{e:HQET-Lagrangian}-\eqref{e:dahqet}, is the energy shift 
which absorbs the $1/a$ divergence of the static energy. If the matching is performed 
in HQET to order $\minv$, it absorbs a $1/a^2$ term.
The non-perturbative determination of these parameters 
has been presented at last years
conference~\cite{Blossier:2010vj} and closely follows the general strategy originally 
discussed in~\cite{HeitgerNJ}. We refer the reader to these papers for any unexplained 
notation and further explanations. What still needs to be emphasised at this point is:
\begin{itemize}
  \item light quarks are simulated with two-flavor non-perturbatively improved Wilson fermions, 
  \item for the static action we use the so-called HYP2 
        discretisation~\cite{DellaMorte:2005yc},%
        \footnote{A final analysis with increased statistics will also take into account 
                  results for HYP1 which are fully compatible.}
  \item in the course of the non-perturbative (NP) matching to QCD in small volume of extent $L_1\sim 0.5\,\fm$, 
        various heavy quark masses have been simulated relativistically at 
        renormalization group invariant (RGI) heavy quark mass $M$ fixed to~\cite{Fritzsch:2010aw}%
        \begin{align} \label{eq:z-values}
                z \equiv L_1 M \in \{ 4,6,7,9,11,13,15,18,21\} \;,  
        \end{align} 
        ranging from slighty above the charm to beyond the bottom quark region.
\end{itemize}
To summarize, the parameters of HQET, $\omega(M,a)$, are known for the masses given in~\eqref{eq:z-values}
and the lattice spacings $a$ corresponding to $\beta\in\{5.2,5.3,5.5\}$ in large volume, c.f. Table~\ref{tab:CLS-ensembles}.

\begin{table}
\vspace{-0.30cm}
\small
\includegraphics*[height=0.5cm, trim= 0  0 0 0 ]{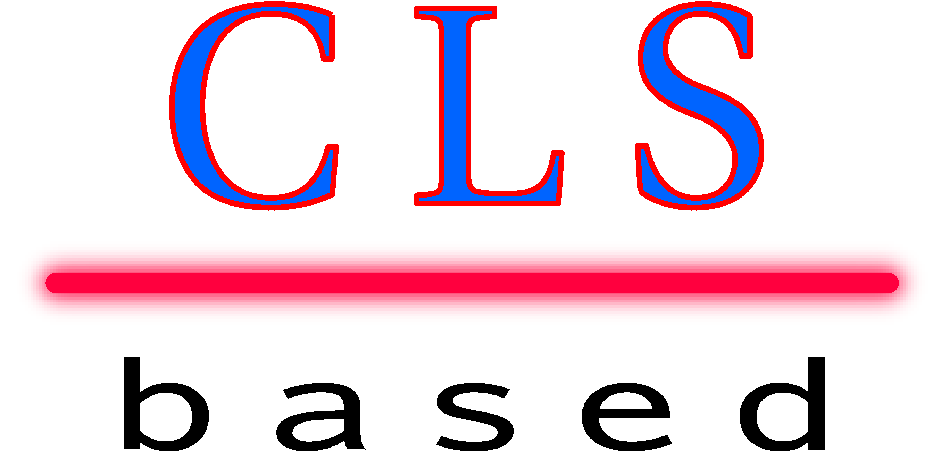}
        \begin{center}
\vskip-1cm
        \begin{tabular}{ccccccccc}\toprule
$\beta$ &   $a$   & $L/a$&$L\mpi$&$\mpi$ & no. of  & separ.     &      \\[-0.25em]  
        & (\fm)   &      &      & (\MeV) & cnfg.s  &  (MD u.)   & label\\\midrule  
 $5.2$  & $0.075$ & $32$ & $4.7$& $386$  &  $800$  &  $8$       &  A4  \\[-0.25em]  
        &         & $32$ & $4.0$& $331$  &  $200$  &  $4$       &  A5  \\           
 $5.3$  & $0.065$ & $32$ & $4.7$& $438$  & $1000$  & $16$       &  E5  \\[-0.25em]  
        &         & $48$ & $4.8$& $312$  &  $500$  &  $8$       &  F6  \\[-0.25em]  
        &         & $48$ & $4.2$& $267$  &  $600$  &  $8$       &  F7  \\           
 $5.5$  & $0.048$ & $48$ & $5.2$& $442$  &  $400$  &  $8$       &  N5  \\[-0.25em]  
        &         & $64$ & $4.2$& $268$  &  $700$  &  $4$       &  O7  \\\bottomrule
        \end{tabular}
        \end{center}
\vspace{-0.4cm}
  \caption{Presently used large volume ensembles of the Coordinated Lattice Simulations (CLS) consortium.}
  \label{tab:CLS-ensembles}
\end{table}

Now we use the effective parameters to compute the $B$ meson mass $\mB$,
the hyperfine mass splitting $m_{\rm B^*}-\mB$ and the pseudo-scalar heavy-light meson
decay constant $\fB$ ($\fBs$ is left for a future application). To first order in 
the $1/\mb$ expansion our main observables are defined by
\begin{align}
                           \mB &= m_{\rm bare} \,+\,E^{\rm stat}\,+\, \omega_{\rm kin} \, E^{\rm kin} \,+\,\omega_{\rm spin} \, E^{\rm spin}\,,  
  \label{eq:mB}\\
  m_{\rm B^*}-\mB &= -\tfrac{4}{3} \omegaspin\,\Espin \;,  
  \label{eq:Deltam}\\
  \ln(a^{3/2}\fB\sqrt{\mB/2}) &= \ln(\zahqet)+ \ln(a^{3/2}p^{\rm stat})+ b^{{\rm stat}}_{\rm A} am_{\rm q} 
  \nonumber \\
                               &{\phantom{=}} + \omega_{\rm kin}  p^{\rm kin} +
  \omega_{\rm spin}  p^{\rm spin} + \cah{1} p^{\rm A^{(1)}} ,
  \label{eq:fB}
\end{align}
while their equivalents at static order are given by choosing 
$\omega= \big( \mhbare^{\rm stat}, \ln[\zastat],  a\castat, 0, 0 \big)$ compared to~\eqref{eq:omegas}. 
The improvement coefficient $b^{{\rm stat}}_{\rm A}$ is determined to one loop in~\cite{Grimbach:2008uy} 
and the HQET energies $\Estat, \Ekin, \Espin$ as well as the HQET hadronic matrix elements 
$p^{\rm stat},p^{\rm kin}, p^{\rm spin}, p^{\rm A^{(1)}}$ depend on the set $(\mpi,a)$, i.e. 
the simulated pion masses $\mpi$ and lattice spacings $a$. 
They have been measured on a subset of ensembles produced within the CLS effort~\cite{CLS} 
with $\nf=2$ flavors of ${\rm O}(a)$-improved Wilson-Clover fermions, obeying 
\begin{align}
     \mpi L &\gtrsim 4.0 \;,  & 
     270\,\MeV &\lesssim \mpi   \lesssim 450\,\MeV \;.
        \label{eq:CLS-bounds}
\end{align}
For this reason we expect finite volume effects to be negligible.
Details of these large volume ensembles including the actual size of the statistics used 
here are summarized in Table \ref{tab:CLS-ensembles}.
\begin{figure}[tb]
\vspace{-0.45cm}
\begin{center}
\includegraphics*[height=3.25cm]{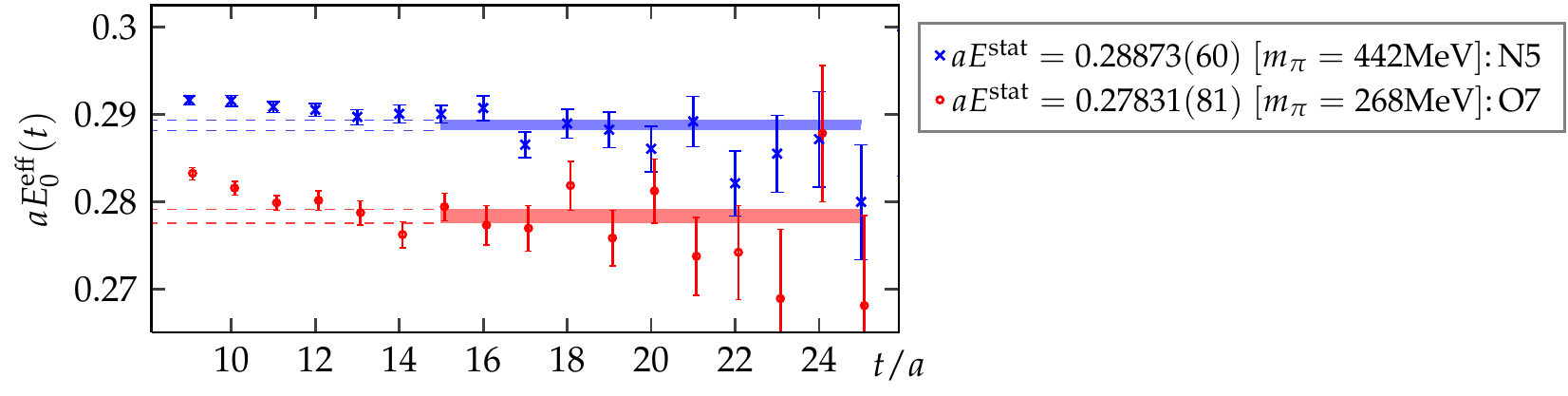}
\end{center}
\vspace{-0.65cm}
\caption{Results for static ground state energies (HYP2) after applying the GEVP method on 2 ensembles. 
}
\label{fig:GEVP-Estat}
\end{figure}
We get HQET energies \& hadronic matrix elements by solving the generalised eigenvalue problem, GEVP, 
\begin{align}
    C(t) v_{n}(t,t_{0}) &= \lambda_{n}(t,t_{0}) C(t_{0}) v_{n}(t,t_{0})\,,
        \label{eq:GEVP}
\end{align}
for an $N\times N$ correlator matrix $C$. The $n^{\rm th}$ state has eigenvalue $\lambda_n$ with eigenvector $v_n$. 
As has been shown in~\cite{BlossierKD}, a systematic expansion of $C$ in HQET is given by
$ C(t) = C^{\rm stat}(t) + {\sum}_{{\rm X}} \omega_{ {\rm X}} C^{ {\rm X}}(t) + {\rm O}(\omega^2)$
in terms of small expansion parameters $\omega_{ {\rm X}}\propto 1/\mb$ with 
${\rm X}\in\{ {\rm kin}, {\rm spin}, {\rm A}^{(1)} \}$, depending on the actual quantity.
The eigenvalue $\lambda_n$ determines the effective energy of the $n^{\rm th}$ state
while the eigenvectors enter the expressions for the matrix elements~\cite{BlossierKD}.
As a variance reduction technique we employ stochastic all-to-all propagators.
The heavy-light interpolating quark bilinears used,
\begin{align} 
O_k(x)   &= \psibar_{\rm h}(x)\gamma_0\gamma_5\psi_{\rm l} {}^{(k)}(x) \,, &
            \psi_{\rm h}(x)       &\text{: static quark field}                     \notag\\[-0.2em]
O_k^*(x) &= \psibar_{\rm l}{}^{(k)}(x)\gamma_0\gamma_5\psi_{\rm h}(x) \,, &             
             \psi_{\rm l}{}^{(k)}\!(x) &= ( 1+a{}^2\,\Delta/10 ){}^{R_k} \psi_{\rm l}(x)
\end{align}
represent different levels of Gaussian smearing \cite{wavef:wupp1} for the light quark field $\psi_{\rm l}{}^{(k)}$,
$k=1,\ldots,N$, with APE smeared links \cite{smear:ape,Basak:2005gi} in the lattice Laplacian $\Delta$.
Numerical experiments have shown that choosing $N=3$ with $R_k\times(a/0.3\fm)^2\in\{1,4,10\}$ 
fixed, gives good results. 
For further details about the general procedure see~\cite{BlossierVZ}.  
Special care has been taken to control the contribution of excited states.
In our analysis we have chosen a time range to extract the plateaux such that the 
corrections to $E^{\rm stat}_{1}$ are small compared to its statistical error; 
we found this to be the case for $t>t_{0}>0.3\,\fm$. Figure~\ref{fig:GEVP-Estat} 
shows two ground state energies obtained in this way. An autocorrelation analysis
has shown that existing data can be considered decorrelated to a sufficient degree.  

We arrange all data sets for a combined jackknife analysis with 100 estimators and compute 
\eqref{eq:mB}-\eqref{eq:fB} which now depend on $(z,\mpi,a)$. For each
quantity we perform a joint, continuum ($a\to 0$) and chiral 
($\mpi\to\mpi^{\rm exp}\equiv m_{\pi^{0}}=135\,\MeV$~\cite{PDG:2010}) 
extrapolation --- $\chi\!+\!$CL in short form.

\section{Results}
\noindent{\bf b-quark mass:}
\begin{figure}[tb]
\vspace{-0.45cm}
\begin{center}
\includegraphics*[height=5cm]{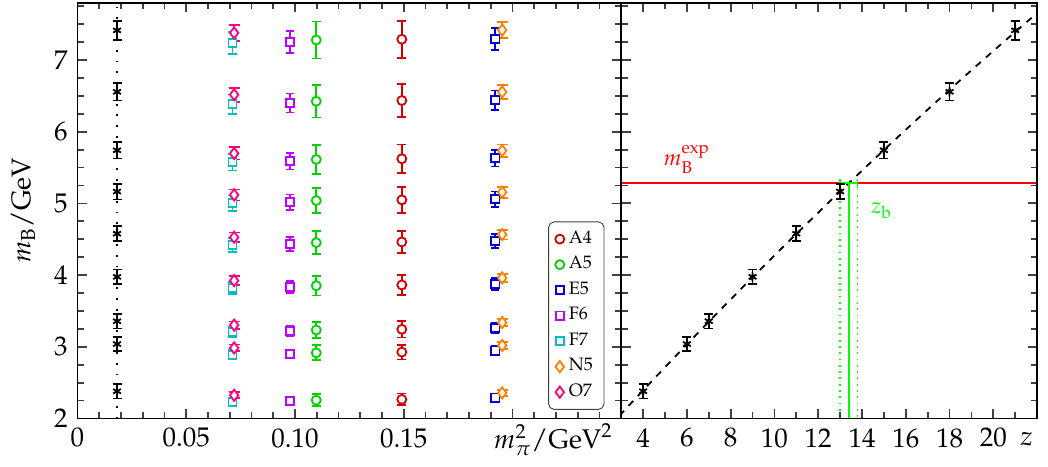}
\end{center}
\vspace{-0.7cm}
\caption{%
{\it Left:} HQET to ${\rm O}(1/\mb)$ data and results of $\chi\!+\!$CL extrap. of $B$ 
meson mass to the physical point for all values of $z$. 
{\it Right:} Dependence $\mB^{\rm HQET}(z)$ in the continuum and its matching to the 
physical $B$ meson. 
}
\label{fig:CL-mB-and-matching}
\end{figure}
We determine $m_{\rm B}$ by combining the HQET parameters and energies using the static
or the O($1/m_{\rm b}$) expressions in eq.~\eqref{eq:mB}. In both cases we then use the
global fit ansatz 
\begin{align}
   m_{\rm B}\left(z,\mpi,a\right) &= B(z)+C\cdot m^2_\pi+C^\prime\cdot m^3_\pi + D\cdot a^2  \,,  
   \label{eqn:fit-mB}
\end{align}
where we either set $C^\prime=0$ or $C^\prime=- 3 \hat{g}^2 / (16 \pi f^2_\pi) $ computed with 
$f_\pi\equiv f_{\pi^{0}}=130.4\,\MeV$~\cite{PDG:2010}
and $\hat{g}=0.51(2)$~\cite{MDonnellan}.
The data points and resulting values of $m_{\rm B}$ at the physical point (from $C^\prime=0$) are shown on
the left of Figure~\ref{fig:CL-mB-and-matching}, while its $z$-dependence is shown on the right.
Knowing the latter allows to match our computations of the $B$ meson mass 
to its physical value,
\begin{align}
       \left. m_{\rm B}(z,\mpi^{\rm exp},0)\right|_{z=z_{\rm b}} &\equiv m_{\rm B}^{\rm exp} \,, &
       \text{taking}\quad
       m_{\rm B}^{\rm exp}&=5279.50\,\MeV\text{~\cite{PDG:2010}}\,. 
        \label{}
\end{align}
As a result we obtain the dimensionless RGI b-quark mass $z_{\rm b}=L_1M_{\rm b}$. 
Taking the recent estimate of $L_1=0.405(18)\,\fm$~\cite{ScaleL1} and applying
the 4-/3-loop running of the coupling/mass in the $\MSbar$ scheme, we obtain 
the b-quark mass to
\begin{align}      
   \left.\overline{m}_{\rm b}(\overline{m}_{\rm b})\right|^{\rm stat}_{\nf=2}\,
              &=4.21(13)_{\rm stat}(3)_a(6)_{z}\,\GeV \;,   
              &\big[ z_{\rm b}^{\rm stat}&=13.46(21)_{\rm stat}(14)_{a}(18)_{z}\big]  \\[0.2em]
   \left.\overline{m}_{\rm b}(\overline{m}_{\rm b})\right|^{\rm HQET}_{\nf=2}
              &=4.23(13)_{\rm stat}(3)_a(6)_{z}\,\GeV \;.   
              &\big[z_{\rm b}^{\rm HQET}&=13.40(22)_{\rm stat}(16)_{a}(18)_{z}\big]  
        \label{}
\end{align}
We explicitly separate the statistical error and the error coming from the scale
setting. 
The last uncertainty is mainly due to the quark mass renormalisation constant 
$Z_{\rm M}$, entering $z$ in QCD~\cite{Fritzsch:2010aw}, i.e. the accuracy to which 
the values of $z$ could be fixed. 
Since the difference of the static and HQET result of $\overline{m}_{\rm b}$ is already
small, we can conclude that the truncation error of the latter $\sim{\rm O}(\Lambda^3/m_{\rm b}^2)$
is negligible compared to the statistical error.
Comparing our new result for example with 
$\overline{m}_{\rm b}(\overline{m}_{\rm b})= 4.163(16)\,\GeV$~\cite{ChetyrkinFV} or
$4.19^{+0.18}_{-0.06}\,\GeV$~\cite{PDG:2010} shows good agreement given our current 
uncertainty. 
Besides an increase in our statistics and the inclusion
of two new ensembles, we are currently improving our estimate of $L_1$ 
which is the dominating source of error in $\mb$ and are reducing the 
error in the lattice spacings.

With the physical value of $z_{\rm b}$ our theory is now uniquely defined and we
can make further predictions to compare with experiment. There are two ways to proceed
from here: a) one can combine the HQET parameters with matrix elements
and/or energies to build new quantities, perform a $\chi\!+\!$CL extrapolation and
a subsequent interpolation to $z_{\rm b}$ {\it or} b) one once interpolates the HQET
parameters to get $\varpi(a)\equiv\omega(z_{\rm b},a)$, use them in new observables
and perform just the $\chi\!+\!$CL extrapolation. 
Doing both serves as a consistency check.

\begin{figure}[tb]
\vspace{-0.45cm}
\begin{center}
\includegraphics*[height=5cm]{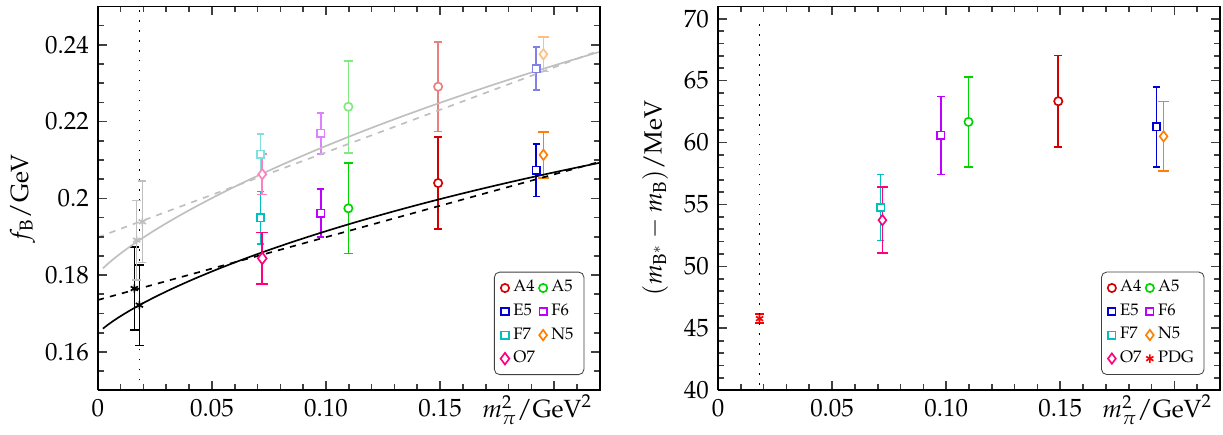}
\end{center}
\vspace{-0.75cm}
\caption{$\chi\!+\!$CL extrapolations to the physical point. {\it Left:} Decay constant $f_B$. {\it Right:} Spin splitting.
}
\label{fig:CL-fB-and-splitting}
\end{figure}

\noindent{\bf Hyperfine/spin splitting:} The leading contribution to this quantity, 
eq.~\eqref{eq:Deltam}, is of ${\rm O}(1/\mb)$ and vanishes in the static limit. Its
intrinsic truncation error thus is ${\rm O}(\Lambda^2/m_{\rm b})$. To construct this 
observable we use $\varpi_{\rm spin}(a)$ which has leading lattice artifacts of 
${\rm O}(a)$. Since this quantity is the one most sensitive to systematic errors, we will
also profit most from a combined analysis with the additional HYP1 data
and the improvements mentioned above. Our results at finite $a$ and simulated pion
masses for the HYP2 action together with the PDG value of $45.78(35)\,\MeV$ at the physical
point are shown in the right panel of Fig.~\ref{fig:CL-fB-and-splitting}.

\noindent{\bf $B$ meson decay constant:} 
Finally we take eq.~\eqref{eq:fB} and rewrite it to extract $f_{\rm B}$ using $\varpi(a)$, 
$\mB^{\rm exp}$ and $a$. Heavy Meson Chiral Perturbation Theory (HM$\chi$PT) predicts a 
dependence 
\begin{align}
   f_{\rm B}\left(\mpi,a\right) &= 
   B\left[ 1 -\tfrac{3}{4} \tfrac{1+3\hat{g}^2}{(4\pi f_\pi)^2}m^2_\pi\ln (m^2_\pi)  \right] + C\cdot m^2_\pi + D\cdot a^2  \,,  
   \label{eqn:fit-fB}
\end{align}
of $\fB$ on $\mpi$, taking $\hat{g}, f_{\pi}$ as in~\eqref{eqn:fit-mB}. The resulting 
extrapolation is shown left in \fig{fig:CL-fB-and-splitting} (solid curve).
If the chiral logarithm is neglected one gets a leading order (LO) fit ansatz with
a linear dependence on $\mpi^2$ (dashed line). The additional, half-transparent data
corresponds to results obtained in the pure static theory. In numbers this reads
\begin{align}
 \text{HM$\chi$PT:}\;\quad\left.{\fB}\right|^{\rm stat}_{\Nf=2} &= 189(6)_{\rm stat}(5)_a\,\MeV\,,  &  
 \text{LO:}        \;\quad\left.{\fB}\right|^{\rm stat}_{\Nf=2} &= 194(6)_{\rm stat}(5)_a\,\MeV\,,  \\
 \boldsymbol{\text{HM$\chi$PT:}\!\quad\left.{\fB}\right|^{\rm HQET}_{\Nf=2}} &\boldsymbol{= 172(6)_{\rm stat}(5)_a\,\MeV}\,,  &  
 \text{LO:}        \!\quad\left.{\fB}\right|^{\rm HQET}_{\Nf=2} &= 176(6)_{\rm stat}(5)_a\,\MeV\,.  
        \label{}
\end{align}
Including the chiral logarithm
of HM$\chi$PT or not, changes the value at the physical point by a small amount only.
At the moment we take the average of the two extrapolations in HQET to ${\rm O}(1/\mb)$,
\begin{align}
   \left.{\fB}\right|^{\rm HQET}_{\; \Nf=2} &= 174(11)(2)\,\MeV\,,  
        \label{}
\end{align}
as the central value and include half of the difference as part of the systematic error.
Note that our estimate of $\fB$ is lower than other estimates presented at this conference,
see~\cite{Lat11:Davies} for a general review.

\section{Conclusions}

We have reported on the status of the ALPHA Collaboration's heavy quark project to extract 
relevant $B$ physics quantities from $\Nf=2$ lattice simulations in the framework of HQET 
expanded to ${\rm O}(1/\mb)$. 
The measurement of HQET energies and matrix elements has been done using the GEVP approach
on ensembles produced by CLS. For the first time we quote results for
$\overline{m}_{\rm b}$ and $\fB$ in the continuum obtained in large volume 
with three lattice spacings and seven pion masses. 
These values are still preliminary since we are confident to decrease uncertainties 
further by using additional data that has not been taken into account yet. 

Once we have further improved the accuracy of our results, we will also determine $\fBs$,
hadronic parameters of the $B-\overline{B}$ mixing or $B \to \pi$ semileptonic form factors
as well as more details of the spectrum of hadrons with a b-flavor. 
As a long term goal we would also like to include the dynamical strange and charm sea quark 
contributions such that the only remaining systematic effect comes from the truncation
of HQET.

\small
\acknowledgments
This work is supported in part by the SFB/TR~9  and grant HE~4517/2-1 of the 
Deutsche Forschungsgemeinschaft and by the European Community through EU 
Contract No.~MRTN-CT-2006-035482, ``FLAVIAnet''. 
We thank F. Bernardoni for discussions on HMChPT and CLS for the joint production 
and use of gauge configurations. 
We thankfully acknowledge the computer resources and support provided by
the John von Neumann Institute for Computing at FZ J\"ulich, of the HLRN 
in Berlin, and at DESY, Zeuthen.

\end{document}